\documentclass[aps,prl,reprint,groupedaddress,showpacs]{revtex4-1}

\usepackage{latexsym,amssymb,amsfonts,currvita, mathrsfs,color,graphicx,amsmath, bm, bbm,fontenc}

\begin{document}

\title{Dynamic clustering and chemotactic collapse of self-phoretic active particles}
\author{Oliver Pohl}
\author{Holger Stark}
\date{\today}

\affiliation{Institut f\"{u}r Theoretische Physik, Technische Universit\"{a}t Berlin, Hardenbergstrasse 36, 10623 Berlin, Germany}

\begin{abstract}
Recent experiments with self-phoretic particles at low concentrations show a pronounced dynamic clustering
[I. Theurkauff \emph{et al.}, Phys.\ Rev.\ Lett.\ \textbf{108}, 268303 (2012)]. We model this situation by taking into
account the translational and rotational diffusiophoretic motion, which the active particles perform in their
self-generated chemical field. Our Brownian dynamics simulations show pronounced dynamic clustering only 
when these two phoretic contributions give rise to competing attractive and repulsive interactions, respectively.
We identify two dynamic clustering states and characterize them by power-law-exponential distributions.
In case of mere attraction a chemotactic collaps occurs directly from the gas-like into the collapsed state,
which we also predict by mapping our Langevin dynamics on the Keller-Segel model for bacterial chemotaxis.
\end{abstract}


\maketitle
The collective motion of self-propelling objects is a most fascinating subject which has been studied in a variety of systems
\cite{Vicsek12,Marchetti13}. At the macroscale, collective patterns occur, for example, in flocks of birds or fish schooles \cite{Cavagna, Bialek_speed, Fishschools} while at the microscopic scale bacterical cells in an aqueous environment generate intricate motional patterns
\cite{Saragostia,Baer,Wensink12}. To understand basic features of structure formation in non-equilibrium, systems with
spherical or circular microswimmers are investigated. A number of theoretical and experimental studies have demonstrated 
that activity of microswimmers alone can result in clustering and phase separation 
\cite{Tailleur08,Marchetti,Marchetti2012,Loewen2012,Baskaran_active,Bechinger,Cates2013,Loewen2013,Stenhammer_Cates} 
due to reduced motility in dense aggregates\ \cite{Tailleur08,Cates2013}.
However, the colloidal density has to be large enough that the characteristic time for a particle to join a cluster becomes comparable 
to its rotational diffusion time needed to dissolve from it \cite{Baskaran_active}.
Other investigations explore the influence of hydrodynamics on collective motion\ \cite{Ishikawa08b,Evans11,Thutupalli11,Fielding12,Alarcon13,Molina13,Zoettl14}.

In experiments with dilute suspensions of self-phoretic active Janus colloids, dynamic clustering has been observed \cite{Bocquet2012, Palacci}.
In this novel non-equilibrium phenomenon, particles constantly join and leave clusters which exhibit a very dynamic shape. Since the
colloids consume a chemical, they create a non-uniform chemical field around themselves. The chemical gradients initiate diffusiophoresis 
\cite{Anderson}, whereby colloids can attract each other as demonstrated in \cite{Palacci}.
The diffusiophoretic mechanism not only provides a novel colloidal interaction, it also serves as a biomimetic version of bacterial
chemotaxis \cite{Berg04}, where cells identify and swim along chemical gradients. Autochemotactic cells typically conglomerate in 
large aggregates or even exhibit a chemotactic collaps \cite{Keller,Brenner,Johannes,Baer}.

Recent theoretical and experimental studies included short-range attraction between active colloids 
and observed clustering at low colloidal densities \cite{Mognetti13,Redner13,Palacci, Schwarz12}.

Ref.\ \cite{Ramaswamy} implements diffusiophoresis for concrete surface properties of self-phoretic
colloids and indentifies various states such as clumping and asters.

The work presented here has very much been inspired by the experiments of the Lyon group\ \cite{Bocquet2012}. The diffusiophoretic interaction
has a translational and orientational contribution. Using Brownian dynamics simulations, we demonstrate that pronounced dynamic clustering 
occurs only when these two contributions give rise to competing attractive and repulsive interactions, respectively. We identify
two dynamic clustering states and characterize them. Otherwise, in case of mere attraction a chemotactic collaps occurs directly from 
the gas-like state before pronounced clusters are able to form. We support this result by mapping our Langevin dynamics on the Keller-Segel 
model for bacterial chemotaxis.

In our model we consider a dilute suspension of $N$ 
self-phoretic colloids confined in a quadratic box.
For example, by
adding the chemical $\mathrm{H_2O_2}$, they become active due to self-electrophoresis \cite{Paxton_self_electrophoresis}
and the colloids move with a swimming velocity $v_0$. It depends on the concentration $c$ of the
chemical that is a control parameter in the experiments \cite{posner2010,posner2011}. We assume that $v_0$ does not change
noticeably due to local inhomogeneities of $c$ and consider it as a system parameter \cite{Bocquet2012,remark_concentration}.
In addition, the active colloids experience diffusiophoretic forces along chemical gradients, which are generated 
by the colloids themselves when they consume the chemical.
Since the colloids are bipolar, a torque
also aligns their directions $\mathbf{e}$ along a chemical gradient. The associated translational and rotational 
diffusiophoretic velocities are given by \cite{Anderson}:
\begin{eqnarray}
\mathbf{v} & = & [ \langle \zeta \rangle \mathbf{1} - \langle  \zeta (3 \mathbf{n} \otimes \mathbf{n} - \mathbf{1}) /2  \rangle ]
\boldsymbol{\nabla} c 
\label{eq.trans} \\
\boldsymbol{\omega} & =  & \frac{9}{4a}  \langle \zeta \mathbf{n} \rangle \times \boldsymbol{\nabla} c .
\label{eq.rot}
\end{eqnarray}
Here the slip-velocity coefficient $\zeta$ depends on the interaction of the chemical with the colloid's surface,
$\langle \ldots \rangle$ is an average over the surface with local normal vector $\mathbf{n}$, and $a$ is the colloid
radius. The quadrupolar term in Eq.\ (\ref{eq.trans}) vanishes for half-coated colloids, whereas the angular velocity 
$\boldsymbol{\omega}$ with $\langle \zeta \mathbf{n} \rangle \propto \mathbf{e}$ is non-zero since the strength of the 
chemical-surface interaction is different for the two halfs of the Janus colloids.
Using $\dot{\mathbf{e}} = \boldsymbol{\omega} \times \mathbf{e}$ and $( \mathbf{e}_i \times \boldsymbol{\nabla} c ) \times \mathbf{e}_i  = (\mathbf{1} - \mathbf{e}_i \otimes \mathbf{e}_i) \boldsymbol{\nabla} c$, we obtain a system of two Langevin
equations describing the two-dimensional position $\mathbf{r}_i$ and direction $\mathbf{e}_i$ of the $i$th colloid
in the overdamped limit:
\begin{eqnarray}
\dot{\mathbf{r}}_i 
& = & v_0 \mathbf{e}_i - \zeta_{\mathrm{tr}} \boldsymbol{\nabla} c(\mathbf{r}_i) + \boldsymbol{\xi}_i,
\label{eq.position}\\
\dot{ \mathbf{e}}_i 
& = & - \zeta_{\mathrm{tr}} (\mathbf{1} - \mathbf{e}_i \otimes \mathbf{e}_i) \boldsymbol{\nabla} 
c(\mathbf{r}_i) + \boldsymbol{\mu}_i \times \mathbf{e}_i  \enspace .
\label{eq.direction}
\end{eqnarray}
Thermal and intrinsic fluctuations enter the equations by translational ($\boldsymbol{\xi}_i$)
and rotational ($\boldsymbol{\mu}_i$) noise with zero mean and correlations 
$\langle \boldsymbol{\xi}_i \otimes \boldsymbol{\xi}_i  \rangle = 2 D_{\mathrm{tr}} \mathbf{1} \delta(t-t')$ and
$\langle \boldsymbol{\mu}_i \otimes \boldsymbol{\mu}_i  \rangle = 2 D_{\mathrm{rot}} \mathbf{1} \delta(t-t')$.
We model the active particles as hard spheres; so, whenever they start to overlap, we separate them along the line
connecting their centers. The chemical field diffuses and has sinks at the positions of the particles since 
they consume the chemical \cite{remark_chemical}:
$
\dot{c}(\mathbf{r}) = D_c \boldsymbol{\nabla}^2 c - k \sum_{i=1}^{N} \delta(\mathbf{r} - \mathbf{r}_i )
\label{eq.cfield} 
$.
Since the chemical diffuses much faster than the colloids move, a stationary density field develops at each instance:
\begin{equation}
c(\mathbf{r}) = c_0 - \frac{k}{4\pi D_c} \sum_{i=1}^{N} \frac{1}{|\mathbf{r} - \mathbf{r}_i |} \enspace .
\label{eq.profile}
\end{equation}
In experiments, active colloids settle on a surface and swim in two dimensions while the chemical field diffuses freely in half-space. 
Implementing zero flux at the surface by image colloids does not change the form of Eq.\ (\ref{eq.profile}). The same holds when we 
integrate the concentration over a layer with thickness twice the colloid radius $a$ to make the relevant concentration field two-dimensional.

Within clusters of active colloids, the concentration field cannot freely diffuse. So, whenever a colloid is surrounded
by six closely packed neighbours, we introduce for it a screened chemical field, $\exp[-(r - \xi ) / \xi] / r$, 
with the screening length $\xi = 2 a ( 1 + \epsilon) $. In our simulations we typically take $\epsilon = 0.3 $ and have checked that
our results do not change when $\varepsilon$ is changed by 50\%.
We rescale time and length by $t_r = 1/(2D_{\mathrm{rot}})$ and $l_r = \sqrt{D_{\mathrm{tr}} / D_{\mathrm{rot}}} 
=  2.33 a$    
respectively.
For historical reasons this is above the experimental value of $1.79 a$ \cite{Bocquet2012}, while the thermal value is $1.15a$. 
Again, the following results do not change drastically, if we adjust the value. The noise intensities for rescaled 
$\boldsymbol{\xi}_i$ and $\boldsymbol{\mu}_i$ become one and we are left with three essential parameters: the Peclet number
$\mathrm{Pe} = v_0 / (2 \sqrt{ D_{\mathrm{tr}}  D_{\mathrm{rot}} } ) $ and reduced translational as well as rotational diffusiophoretic
parameters
$ \zeta_{\mathrm{tr}} k / (8\pi D_{\mathrm{tr}}D_c  )  \rightarrow \zeta_{\mathrm{tr}} $ and
$ \zeta_{\mathrm{rot}} k / (8\pi D_c \sqrt{ D_{\mathrm{tr}} D_{\mathrm{rot}}} )  \rightarrow \zeta_{\mathrm{rot}} $,
respectively, for which we keep the same symbols. 

Note that in our model each active 
colloid creates a chemical sink.  So, for $\zeta_{\mathrm{tr}} > 0$ and  $\zeta_{\mathrm{rot}} > 0 $ it pulls and rotates other colloids towards itself, which means an effective attraction. 
At boundaries of the simulation box, we let colloids move with 
a randomly chosen direction back into the box. Following experiments \cite{Bocquet2012}, we investigate dilute systems 
with an area fraction of $\sigma = 5 \%$ and simulate a total of 800 particles. We have checked that a larger number does 
not change our results as long as the area fraction is maintained. Finally, we choose the swimming velocity $v_0$ in the 
range $2-4.5 \mathrm{\mu m /s}$ as in experiments \cite{Bocquet2012} corresponding to Peclet numbers $\mathrm{Pe} = 10 - 22$.

\begin{figure}
\includegraphics[height=0.4\textwidth]{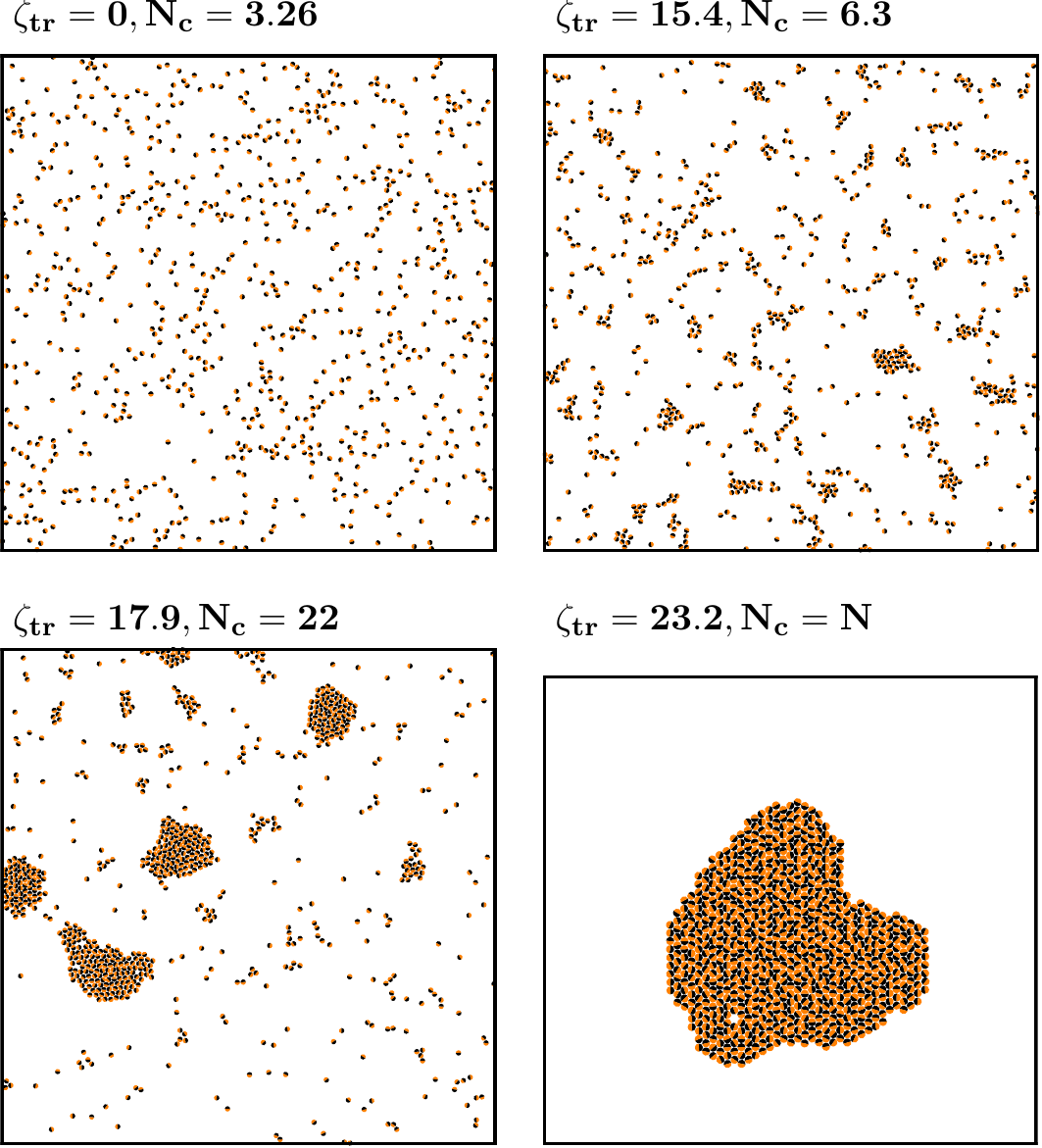}
\caption{
Snapshots of colloid configurations for increasing $\zeta_{\mathrm{trans}}$ at $\zeta_{\mathrm{rot}}=-0.38$ and $\mathrm{Pe}=19$.
Top left: Gas-like state, top right: dynamic clustering 1, bottom left: dynamic clustering 2, bottom right: collapsed state.
$N_c$ is the mean cluster size.
}
\label{fig:snap}
\end{figure}

\begin{figure}
\includegraphics[height=0.313\textwidth]{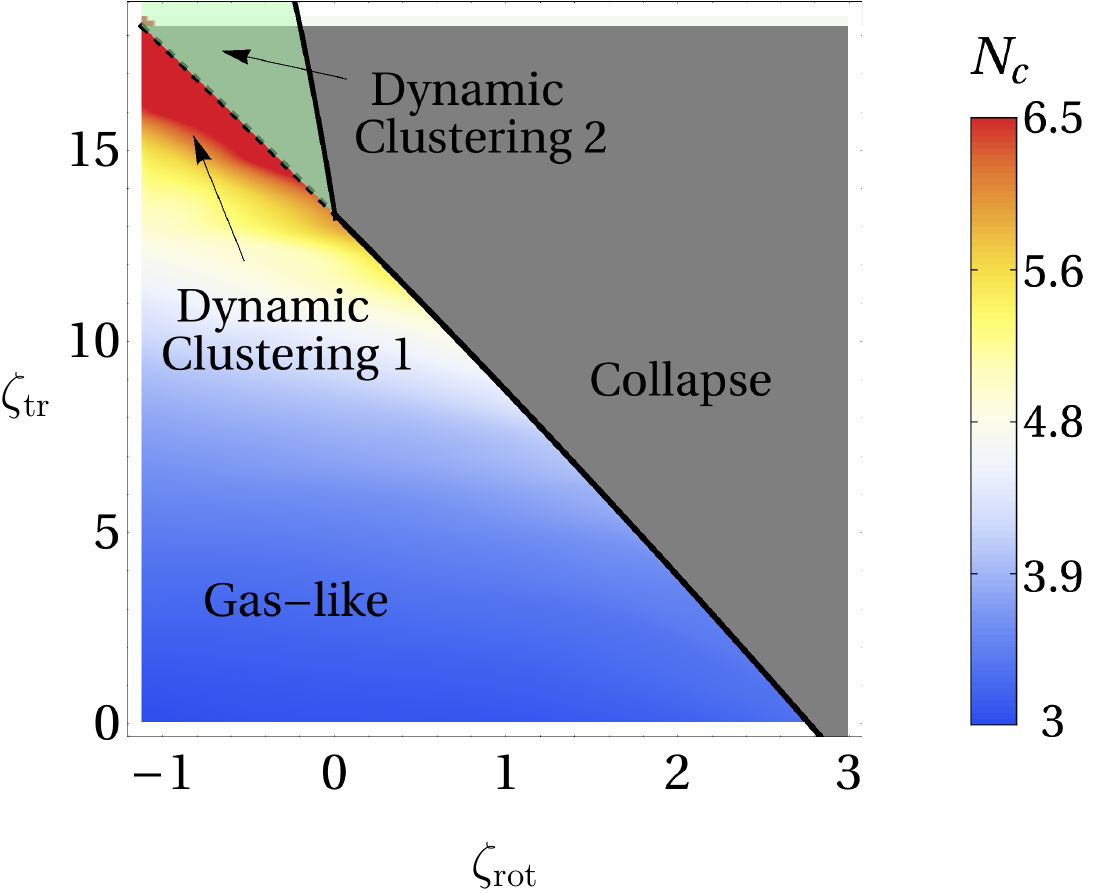}
\caption{
State diagram: $\zeta_{\mathrm{tr}}$ versus $\zeta_{\mathrm{rot}}$ at $\mathrm{Pe} = 19$. The mean cluster size $N_c$
is color coded.  
}
\label{fig:phase_rot}
\end{figure}

In Fig.\ \ref{fig:snap} we illustrate typical particle configurations for increasing translational diffusiophoretic parameter 
$\zeta_{\mathrm{tr}}$. At $\zeta_{\mathrm{tr}} = 0$ (upper left) the active colloids hardly cluster since there is no phoretic 
attraction between them. They assume a gas-like phase with a mean cluster size $N_c$ close to 3.
To determine $N_c$, we define a cluster as an assembly of more than two colloids and average over many snapshots
at different times. Thus, by definition $N_c \ge 3$. In contrast, at large $\zeta_{\mathrm{tr}}$ (bottom right) the system 
collapses into a single large cluster similar to the chemotactic collaps which occurs in bacterial systems \cite{Keller, Blowup_Jaeger}.
In between, we observe dynamic clustering (top right, bottom left). Motile clusters form that strongly fluctuate in shape 
and size and ultimately dissolve again (see movie 1 and 2 in the supplemental material). 
In Fig.\ \ref{fig:phase_rot} we plot a state diagram in the phoretic parameters
$\zeta_{\mathrm{tr}}$, $\zeta_{\mathrm{rot}}$, where we also color-code the mean cluster size $N_c$.
At $\zeta_{\mathrm{rot}} = 0$ the maximal dynamic cluster size just before the collapse is $N_c \approx 5$. When we
turn on rotational diffusiophoresis with $\zeta_{\mathrm{rot}} >  0$, the swimming direction $\mathbf{e}$ of a free 
active colloid points towards a cluster (chemical sink) which further supports the formation of one cluster. This 
explains the fact that in Fig.\ \ref{fig:phase_rot} the collapse occurs for smaller $\zeta_{\mathrm{tr}}$ when $\zeta_{\mathrm{rot}}$
increases. The mean cluster size just before the collapse decreases and dynamic clustering is hardly visible. In contrast,
at $\zeta_{\mathrm{rot}} < 0$ active particles rotate away from chemical sinks and thus an effective repulsion is
introduced. Once active colloids join a cluster, their swimming direction rotates outwards and the colloids can leave the
cluster again if the translational phoretic attraction is not too large. This balance of effective phoretic attraction and repulsion
is the cause for pronounced dynamical clustering with large cluster sizes.
Interestingly, the state diagram in Fig.\ \ref{fig:phase_rot} indicates two clustering states: 
one where cluster sizes up to 6.5 are observed (see snaphot top, right in Fig.\ \ref{fig:snap}) 
and a second clustering state where much larger dynamic clusters occur (see snaphot bottom, left
in Fig.\ \ref{fig:snap}). We will characterize these two states in the next paragraph. In a chemotaxis model for bacteria, the chemotactic collapse has been rationalized by Keller and Segel \cite{Keller}.
Below, we will demonstrate that our model
can be mapped on the Keller-Segel equation, which qualitatively explains the 
nearly straight transition line in the state diagram 
between the gas-like and the collapsed state.

\begin{figure}
\includegraphics[height=0.29\textwidth]{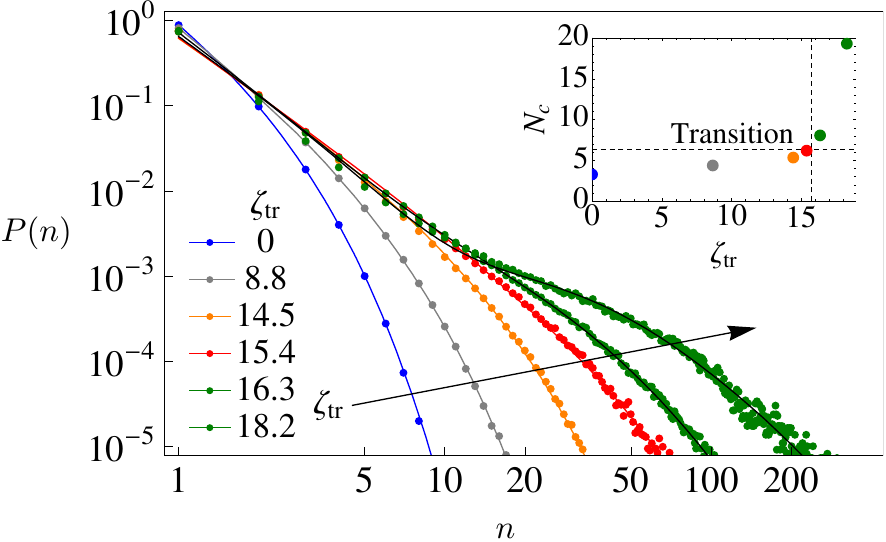}
\caption{
Cluster size distributions $P(n)$ for increasing translational phoretic parameter $\zeta_{\mathrm{tr}}$
at $\mathrm{Pe}=19$ and $\zeta_{\mathrm{rot}} = -0.38$. $\zeta_{\mathrm{tr}}$ assumes the
values 0, 8.8, 14.5, 15.4, 16.3 and 18.2. The transition between dynamic clustering states
1 and 2 occurs between the red and green curves. Inset: mean cluster size $N_c$ versus $\zeta_{\mathrm{tr}}$. The transition
is indicated.
}
\label{fig:distribution}
\end{figure}

To further quantify the two dynamic clustering states 1 and 2, we determine the cluster-size distribution $P(n)$.
In Fig.\ \ref{fig:distribution}
we plot it for fixed rotational phoretic parameter $\zeta_{\mathrm{rot}}=-0.38$ and increasing translational coupling $\zeta_{\mathrm{tr}}$. 
For pure steric interaction ($\zeta_{\mathrm{tr}} = 0$, blue curve), an exponential decay is predominant.
Closer to the transition line between dynamic clustering states 1 and 2 in Fig.\ \ref{fig:phase_rot}, the distribution follows a power law at small $n$, before it falls off exponentially (orange and red curves in Fig.\ \ref{fig:distribution}).
Indeed, we can fit our results by $P(n) = c_1 n^{-\beta} \exp(-n/n_0)$ with exponent $\beta_1 = 2.1 \pm 0.1$, which gradually decreases
for more negative $\zeta_{\mathrm{rot}}$.
With further increase of $\zeta_{\mathrm{tr}}$ we observe that the distribution $P(n)$ develops an inflection
point (green curves) and we have to use a sum of two power-law-exponential curves to fit our distributions,
$p(n)=c_1  n^{-\beta_1}\exp(-n/n_1) + c_2 n^{-\beta_2}\exp(-n/n_2)$ with $\beta_1 = 2.1 \pm 0.2$ and
$\beta_2 \approx 1.5$.  
This defines the dynamic clustering state 2, where very large clusters coexist 
with smaller ones and individual particles. The transition in the cluster-size distribution is observed
for all negative $\zeta_{\mathrm{rot}}$. Typically, the mean cluster size $N_c$ increases strongly
in state 2 as indicated in the inset of Fig.\ \ref{fig:distribution}. 
In contrast, for $\zeta_{\mathrm{rot}} > 0$ the system exhibits the collapse to a single cluster before 
large dynamic clusters can appear.

Similar forms of the cluster-size distributions including the transition indicated by the occurrence
of an inflection point has been observed in experiments on gliding bacteria\ \cite{Baer}.
However, in this work bacterial density was varied in a system with pure hard-core interactions
causing nematic alignment.
In contrast, in our case we vary the strength of the diffusiophoretic coupling
ultimately leading to
the collapsed state which has not been observed in Ref.\ \cite{Baer}. 

\begin{figure}
\includegraphics[height=0.291\textwidth]{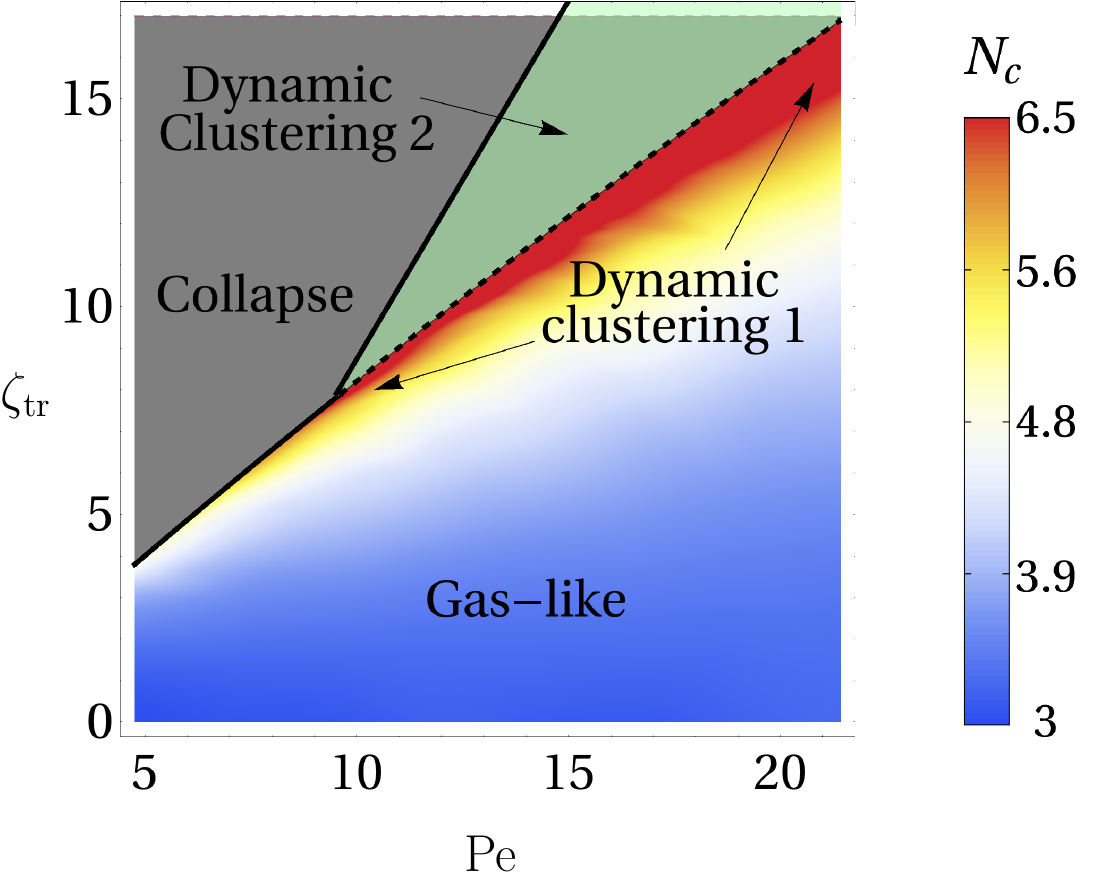}
\caption{
State diagram plotting mean cluster size $N_c$ against Peclet number Pe and chemotactic control parameter $\zeta_{\mathrm{tr}}$.
$\zeta_{\mathrm{rot}} = -0.38 $.
The colour code indicates the mean number of particles, $N_c$, in a dynamic cluster. 
}
\label{fig:phase}
\end{figure}

The experiments with diffusiophoretic coupling showed 
a linear scaling of the mean cluster size with $\mathrm{Pe}$:
$N_c \sim \mathrm{Pe}$ \cite{Bocquet2012}. It appears counterintuitive that at low area fractions of $\sigma = 5 \%$
faster colloids generate larger clusters and indeed the simulations of our model show the contrary behavior. 
In Fig.\ \ref{fig:phase} we plot a state diagram
in $\zeta_{\mathrm{tr}}$ versus $\mathrm{Pe}$ while $\zeta_{\mathrm{rot}}$ is kept constant. Clearly,
for constant $\zeta_{\mathrm{tr}}$ large clusters disappear with increasing $\mathrm{Pe}$. However,
large activity $\mathrm{Pe}$ is necessary for observing dynamic clustering. At small $\mathrm{Pe}$
only small cluster sizes occur with increasing $\zeta_{\mathrm{tr}}$, while at sufficienty large $\mathrm{Pe}$
we observe both dynamic clustering states.

\begin{figure}
\includegraphics[height=0.28\textwidth]{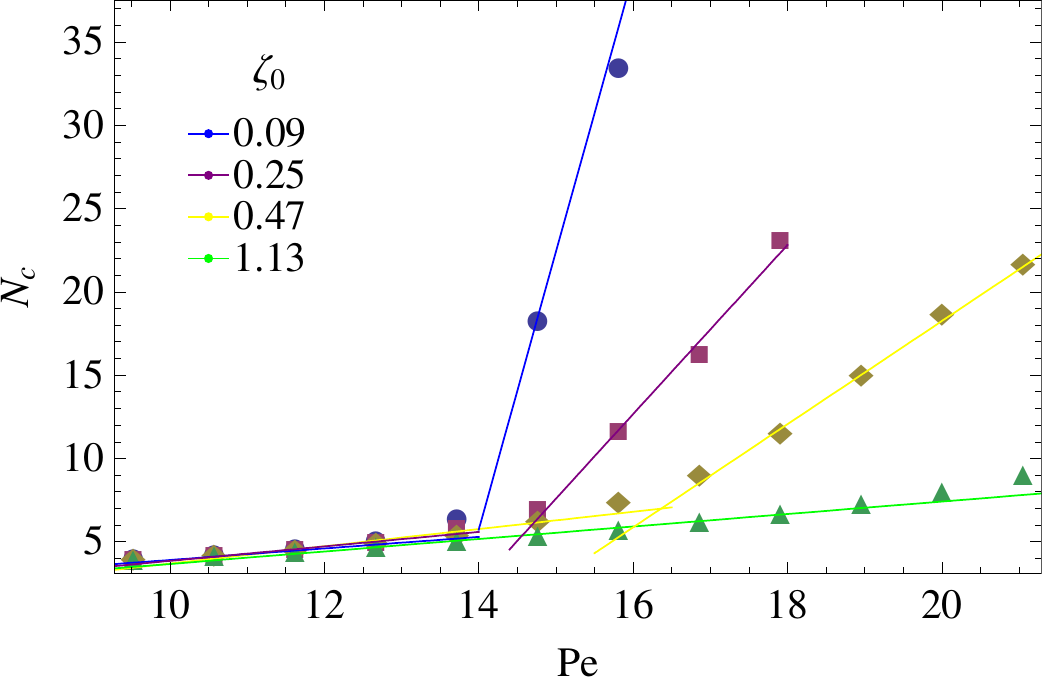}
\caption{Mean cluster size $N_c$ versus $\mathrm{Pe}$ for different lines in the full
parameter space defined via a parametrization with $x\in [0,1]$. We vary $\mathrm{Pe}$ as 
in experiments of \cite{Bocquet2012}, $\mathrm{Pe} = 9.5 + 11.5 x$, choose $\zeta_{\mathrm{tr}} = 4.8 + 16.6 x$ and
$ \zeta_{\mathrm{rot}} = -0.16 - \zeta_0 x$, where the parameter $\zeta_0$ defines the different
graphs. The transition between clustering states 1 and 2 roughly occurs at the intersection
of the two straight lines.
}
\label{fig:lines}
\end{figure}

In the experiments of Ref.\ \cite{Bocquet2012} colloidal activity increases with the concentration $c_0$ of the 
activating chemical $\mathrm{H}_2\mathrm{O}_2$. According to \cite{posner2010}, the swimming velocity
scales as $c_0\sim v_0 \propto \mathrm{Pe}$. 
The concentration $c$ also couples to our reparametrized diffusiophoretic parameters $\zeta_{\mathrm{tr}}$ and $\zeta_{\mathrm{rot}}$ 
through the reaction rate $k$ introduced in Eq.\ (\ref{eq.profile}). In the following, we assume Michaelis-Menten kinetics for the reaction
rate  in the linear regime well before saturation, $k\sim c_0$, and thus find $\zeta_{\mathrm{tr}}\sim c_0 \sim \zeta_{\mathrm{rot}}$.
So, the linear dependence of the parameters $\mathrm{Pe}$, $\zeta_{\mathrm{tr}}$, and $\zeta_{\mathrm{rot}}$ on $c_0$ defines
a line in the parameter space. In this space the dynamic clustering states 1 and 2 are separated by a plane. We choose different 
lines which always hit the transition plane and plot in Fig.\ \ref{fig:lines} the mean cluster size $N_c$
versus $\mathrm{Pe}$ along the lines. The blue and purple curves show the strong increase of $N_c$ when the clustering state\ 2
is entered, since the respective lines hit the transition plane under angles closer to $90^{\circ}$. Making this angle smaller, the increase is more modest.
In particular, the green graph shows an almost linear increase of $N_c$ in the $\mathrm{Pe}$ range from 10 to 20. 
So, Fig.\ \ref{fig:lines} 
demonstrates that the relation between $N_c$ and swimming velocity does not have a simple dependence.

To gain some more insight into the transition to the collapsed state, we write down a Smoluchowski equation for the
full spatial and orientational probability distribution. We determine its orientational moments and derive from them
an equation for the spatial colloidal density $P(\mathbf{r},t)$ coupled to the chemical field $c(\mathbf{r},t)$,
similar to the approach in Ref.\ \cite{Golestanian}:
\begin{align}
\dot{P} &=  \zeta_{\mathrm{eff}} \boldsymbol{\nabla} \cdot (P \boldsymbol{\nabla} c )  + D_{\mathrm{eff}} \boldsymbol{\nabla}^2 P 
+ \mathrm{O}((\boldsymbol{\nabla} c)^2).
\label{eq.eff}
\end{align}
Details of the derivation are given in the supplemental material. Equation (\ref{eq.eff}) is reminiscent of
one relation of the Keller-Segel model \cite{Keller}
used to describe the chemotaxis of bacteria but here with effective chemotactic and diffusion constants:
$\zeta_{\mathrm{eff}} = \zeta_{\mathrm{tr}} + \frac{\zeta_{\mathrm{rot}}v_0}{2D_{\mathrm{rot}}}$ and $D_{\mathrm{eff}}
= D_{\mathrm{tr}} + \frac{v_0^2}{2D_{\mathrm{rot}}}$. In two dimensions the Keller-Segel equation exhibits an instability of the
uniform state towards a chemotactic collaps when its parameters satisfy $\frac{\zeta_{\mathrm{eff}}k \sigma}{D_c D_{\mathrm{eff}}} > b$,
where $\sigma$ is area fraction of the colloids and $b$ a constant which depends on the geometry of the system 
\cite{Blowup_Jaeger, Blowup_96}). In our unitless parameters the condition becomes
$\frac{ 8\pi\sigma(  \zeta_{\mathrm{tr}} + \zeta_{\mathrm{rot}}\mathrm{Pe})}{1 + 2\mathrm{Pe}^2} > b$.  For constant $\mathrm{Pe}$, 
this agrees nicely with the
nearly straight transition line in Fig.\ \ref{fig:phase_rot}  between the gas-like and the collapsed state at $\zeta_{\mathrm{rot}} > 0$ which extents 
to the line separating clustering states 1 and 2. However, the condition does not reproduce the straight transition line in Fig.\ \ref{fig:phase}.
The reason is that at $\zeta_{\mathrm{rot}} < 0$ pronounced clustering occurs before the collapse takes place. Clustering, however, cannot be
described by the Keller-Segel equation since we neglect direct two-particle interactions during its derivation. To test the collapse condition in regions
where dynamic clustering is absent, we determine the transition line $\zeta_{\mathrm{tr}}$ versus $\mathrm{Pe}$ for positive 
$\zeta_{\mathrm{rot}}$ and indeed find a quadratic dependence in $\mathrm{Pe}$ as Fig. 1 in the supplemental material shows.
The inset reproduces the predicted linear dependence between $\zeta_{\mathrm{rot}}$ and $\mathrm{Pe}$ for $\zeta_{\mathrm{tr}} = 0$.
On the other hand, the straight transition line between clustering states 1 and 2 in Fig.\ \ref{fig:phase} follows directly from 
Eq.\ (\ref{eq.position}). At $\zeta_{\mathrm{rot}} < 0$, the particles at the rim of a cluster point away from it. So, large clusters form
when swimming away from the cluster and attraction to the cluster by translational diffusiophoresis balance each other,
which means $v_0 \propto \zeta_{\mathrm{tr}} |\boldsymbol{\nabla} c|$.

Self-phoretic active colloids mediate diffusiophoretic interactions between each other. They consist of a translational and an
orientational part which together with the active swimming can act either attractively or repulsively. When they are both attractive,
the colloids show a transition from the gas-like to a collapsed state reminiscent of a chemotactic collapse in bacterial systems
as our mapping on the Keller-Segel model demonstrates. When translational and rotational diffusiophoresis generate
counteracting attraction  and repulsion, two dynamic clustering states with characteristic cluster size distributions are stabilized
very similar to the dynamic clustering observed in the experiments of \cite{Bocquet2012}. 

The present system mimics chemotaxis in bacterial colonies without relying on a complex signalling pathway necessary in cells. 
Thereby, it may help to explore chemotactic structure formation and design novel dynamic patterns in bacterial colonies \cite{Saragostia}.

\acknowledgments
We thank L. Bocquet, C. Prohm, C. Valeriani, and K. Wolff for helpful discussions and acknowledge funding from the DFG within the research 
training group GRK 1558. This research was supported in part by the National Science Foundation under Grant No. NSF PHY11-25915.
H.S. thanks the Kavli Institute for Theoretical Physics for hospitality and financial support.



\begin{thebibliography}{10}
\bibitem{remark_chemical}
For simplicity, we just consider one chemical field and do not distinguish between reactants and products
of the reaction catalyzed by the active particle.
\end{thebibliography}
\end{document}